# Crossover from a pseudogap state to a superconducting state

Cao Tian-De （曹天德）

*Department of Physics, Nanjing University of Information Science & Technology, Nanjing 210044, China*

**Abstract**

On the basis of our calculation we deduce that the particular electronic structure of cuprate superconductors confines Cooper pairs to be firstly formed in the antinodal region which is far from the Fermi surface, and these pairs are incoherent and result in the pseudogap state. With the change of doping or temperature, some pairs are formed in the nodal region which locates the Fermi surface, and these pairs are coherent and lead to superconductivity. Thus the coexistence of the pseudogap and the superconducting gap is explained when the two kinds of gaps are not all on the Fermi surface. It is also shown that the symmetry of the pseudogap and the superconducting gap are determined by the electronic structure, and non-s wave symmetry gap favors the high-temperature superconductivity. Why the high-temperature superconductivity occurs in the metal region near the Mott metal-insulator transition is also explained.

PACC: 7400, 7420M, 7490

## 1. Introduction

The pseudogap is one of the most pervasive phenomena of high temperature superconductors [1, 2]. There exist two main theoretical scenarios for the explanation of the pseudogap. One is based upon the model of Cooper pairs formation already above the critical temperature of superconducting transition [3, 4], while another assumes that the appearance of the pseudogap state is due to the other orders [5, 6, 7]. Our calculation show that the pseudogap is due to the preformed pairs because the main features of the Fermi arc, the pseudogap and the superconductivity observed in experiments can be explained self-consistently with this idea.

## 2. Calculation

The pseudogap and the high-temperature superconductivity should be due to the electronic origin, thus we consider the affective model

$$H = \sum_{l,l',\sigma} t_{ll'} d^+_{l\sigma} d_{l'\sigma} + U \sum_l n_{l\sigma} n_{l\bar{\sigma}} + \frac{1}{4} \sum_{l,l',\sigma,\sigma'} V_{ll'} n_{l\sigma} n_{l'\sigma'}$$

$$- \sum_{l,l'} J_{ll'} \hat{S}_{lz} \hat{S}_{l'z} \qquad (1)$$

in the $CuO_2$ planes of high-temperature cuprate superconductors. The on-site interaction should be $U > 0$ for a real model. The third and fourth terms have been considered because the on-site interaction $U$ should be not too large for the doped cuprate. Other models similar to this have been solved with many technologies in literatures, however, to show the effects of the electron correlation, we use the charge operator and the spin operator

$$\hat{\rho}(q) = \frac{1}{2} \sum_{k,\sigma} d^+_{k+q\sigma} d_{k\sigma} \qquad (2)$$

$$\hat{S}(q) = \frac{1}{2} \sum_{k,\sigma} \sigma d^+_{k+q\sigma} d_{k\sigma} \qquad (3)$$

to arrive at the Hamiltonian

$$H = \sum_{k,\sigma} \xi_k d^+_{k\sigma} d_{k\sigma} + \sum_q V(q) \hat{\rho}(q) \hat{\rho}(-q)$$

$$- \sum_q J(q) \hat{S}_z(q) \hat{S}_z(-q) \qquad (4)$$

in the wave vector space. Where $V(q) = U + V_0(q)$ and $J(q) = U + J_0(q)$. We have denoted wave vector $\vec{k}$ as $k$, $k \equiv \vec{k}$. The relations $V(-q) = V(q)$ and $J(-q) = J(q)$ will be used below.

We define these functions

$$G(k\sigma, \tau - \tau') = - <T_\tau d_{k\sigma}(\tau) d^+_{k\sigma}(\tau')> \qquad (5)$$

$$F^+(k\sigma, \tau - \tau') = <T_\tau d^+_{k\sigma}(\tau) d^+_{\bar{k}\bar{\sigma}}(\tau')> \qquad (6)$$

$$F(k\sigma, \tau - \tau') = <T_\tau d_{\bar{k}\bar{\sigma}}(\tau) d_{k\sigma}(\tau')> \qquad (7)$$

and establish their dynamic equations, the two-particle Green's functions appear in these equations. Again we establish the dynamic equations of $<T_\tau \hat{S}(q) d_{k+q\sigma} d^+_{k\sigma}(\tau')>$ and $<T_\tau \hat{\rho}(q) d_{k+q\sigma} d^+_{k\sigma}(\tau')>$, and we use the cut-off approximation, then we obtain

$$[-i\omega_n - \xi_k + \sum_q \frac{P(k,q,\sigma)}{i\omega_n + \xi_{k+q}}] F^+(k\sigma, i\omega_n)$$

$$= \Delta^+_{(-)}(k, \sigma, i\omega_n) G(\bar{k}\bar{\sigma}, i\omega_n) \qquad (8)$$

and

$$[i\omega_n - \xi_{\bar{k}} - \sum_q \frac{P(\bar{k},q,\bar{\sigma})}{i\omega_n - \xi_{k+q}}] G(\bar{k}\bar{\sigma}, i\omega_n)$$

$$= 1 + \frac{V(0) <\hat{\rho}(0)>}{i\omega_n - \xi_k} + \Delta_{(+)}(k, \sigma, i\omega_n) F^+(k\sigma, i\omega_n) \quad (9)$$

where

$$\Delta_{(\pm)}(k, i\omega_n)$$

$$= \sum_q \frac{1}{2} \frac{\xi_{k+q} - \xi_k}{-i\omega_n \pm \xi_{k+q}} [J(q) + V(q)] F(k+q, \tau = 0)$$

$$\qquad (10)$$

$$\Delta^+_{(\pm)}(k, i\omega_n)$$

$$= \sum_q \frac{1}{2} \frac{\xi_{k+q} - \xi_k}{-i\omega_n \pm \xi_{k+q}} [J(q) + V(q)] F^+(k+q, \tau = 0)$$

$$\qquad (11)$$

$$P(k,q,\sigma)$$

$$= \frac{1}{2} (\xi_k - \xi_{k+q})(J_0(q) - V_0(q)) G(k+q\sigma, \tau = 0)$$

$$- 2\sigma V(-q) <\hat{\rho}(-q) \hat{S}(q)> J(q)$$

$$+V(-q)<\hat{\rho}(-q)\hat{\rho}(q)>V(q)$$

$$+J(-q)<\hat{S}(-q)\hat{S}(q)>J(q) \qquad (12)$$

Some constant numbers have been absorbed into the chemical potential. $S_{lz}$ is the spin at each site and is zero for non-ferromagnetism. The function $P(k,q,\sigma)$ is called the correlation strength function. This function does depend on the spin index $\sigma$, but the spin dependences could be neglected when we discuss the pairing temperature in the non-magnetic state. This is not to say that the spin dependence of other quantities could be neglected. Because each function $f(k)$ has the relation $f(\bar{k}) \equiv f(-k) = f(k)$, we obtain

$$[-i\omega_n - \xi_k + R_{(+)}(k,i\omega_n) + \frac{\Delta_{(+)}(k,i\omega_n)\Delta^+_{(-)}(k,i\omega_n)}{i\omega_n - \tilde{\xi}_k - R_{(-)}(k,i\omega_n)}]F^+(k,i\omega_n)$$

$$= \frac{\Delta^+_{(-)}(k,i\omega_n)}{i\omega_n - \xi_k - R_{(-)}(k,i\omega_n)}(1 - \frac{V(0)<\hat{\rho}(0)>}{-i\omega_n + \xi_k}) \qquad (13)$$

where

$$R_{(\pm)}(k,i\omega_n) = \sum_q \frac{P(k,q)}{i\omega_n \pm \xi_{k+q}} \qquad (14)$$

$F^+(k,i\omega_n)$ can be found with Eq.(13). When we use $F^+(k,\tau=0) = \frac{1}{\beta}\sum_n F^+(k,i\omega_n)$ to get $F^+(k,\tau=0)$, we need do contour integral, while the integral value is determined by the poles $E_k^{(\pm)}$ of $F^+(k,z)$ near the Fermi surface, thus we obtain the equation

$$\Delta^+_{(\pm)}(k)$$

$$= \sum_q [J(q) + V(q)](\xi_k - \xi_{k+q})n_F(\pm\xi_{k+q})$$

$$\cdot [\frac{\lambda(k,E^{(+)}_{k+q})n_F(E^{(+)}_{k+q})}{\Omega^{(+)}(E^{(+)}_{k+q}) - \Omega^{(-)}(E^{(+)}_{k+q})}$$

$$- \frac{\lambda(k,E^{(-)}_{k+q})n_F(E^{(-)}_{k+q})}{\Omega^{(+)}(E^{(-)}_{k+q}) - \Omega^{(-)}(E^{(-)}_{k+q})}] \cdot \Delta^+_{(-)}(k+q) \qquad (15)$$

where

$$\lambda(k,\omega) = (1 + \frac{V(0)<\hat{\rho}(0)>}{\omega - \xi_k}) \qquad (16)$$

$$\Omega^{(\pm)}(\omega) = \frac{1}{2}\{R_{(-)}(k,\omega) + R_{(+)}(k,\omega) \pm \sqrt{[2\xi_k + R_{(-)}(k\omega) - R_{(+)}(k\omega)]^2 + 4\Delta^+_{(-)}(k\omega)\Delta_{(+)}(k\omega)}\}$$

$$(17)$$

where $E_k^{(\pm)}$ are determined by the equation

$$E_k^{(\pm)} = \Omega^{(\pm)}(E_k^{(\pm)}) \qquad (18)$$

In a similar way, we can find the equations of the function $F(k)$ and $\Delta_{(\pm)}(k)$, while some results could be found with Eq.(15).

### 3. Calculation-based deduction

The pairing temperature could be $T_{pair} > 0$ K and can be found with Eq. (15) for $\Delta^+_{(-)}\Delta_{(+)} \to 0$. This conclusion can be understood because the coefficient of $\Delta^+_{(-)}(k+q)$ in Eq. (15) is positive. However, the values of $T_{pair}$ are determined by the parameters $t_{ll'}$, $U$, $V_{ll'}$ and $J_{ll'}$. Because all factors intending to increase $\Delta_{(\pm)}^{+}$ will increase $T_{pair}$, $T_{pair}$ should be evaluated with $\Delta_{(-)}^{+}\Delta_{(+)} =$ and $\neq 0$.

We find that $E_k^{(+)}$ expresses one of the energy bands with Eq.(9). $E_k^{(\pm)}$ could be found by successive iteration, $E_k^{(\pm,1)} = \Omega^{(\pm)}(\pm\tilde{\xi}_k)$, $E_k^{(\pm,2)} = \Omega^{(\pm)}(E_k^{(\pm,1)})$ … $E_k^{(\pm,n)} = \Omega^{(\pm)}(E_k^{(\pm,n-1)})$. Because $\Omega^{(+)} - \Omega^{(-)} \to 0$ for $E_k^{(+)} - E_k^{(-)} \to 0$ (when $\Delta_{(-)}^{+}\Delta_{(+)} = 0$), to get a higher pairing temperature, Eq.(16) shows that $E_k^{(+)} - E_k^{(-)}$ should tend to zero in as many as possible points near or on the Fermi surface. Of course, the values of $E_k^{(+)} - E_k^{(-)}$ are dominated by the correlation strength function $P$, while $P$ varies with doping and temperature, thus $E_k^{(+)} - E_k^{(-)}$ varies with doping and temperature, and this results in the change of the pairing temperature.

Because the Fermi arc of the cuprate superconductors has been observed in many experiments, thus $E_{k_F}^{(+)} = 0$ in a reasonable model is allowed in the nodal region of the Brillouin zone. Because $E_{k_F}^{(+)} = 0$ is met if $E_{k_F}^{(-)} = 0$ with Eq. (18), thus $E_{k_F}^{(+)} - E_{k_F}^{(-)} = 0$ is allowed in the nodal region of the Brillouin zone. The Fermi segment in the antinode has not been observed in experiments, thus $E_k^{(+)} < 0$ in the antinodal region. This leads us to assume that the energy difference $E_{k_F}^{(+)} - E_{k_F}^{(-)}$ near the node for $\Delta_{(-)}^{+}\Delta_{(+)} = 0$ is zero while $E_k^{(+)} - E_k^{(-)} \neq 0$ in antinode as shown in Fig. 1 & 2. Therefore, to get a higher pairing temperature, $\Delta_{(-)}^{+}\Delta_{(+)}$ (for $T < T_{pair}$ or $T \to T_{pair} - 0$) should be zero around the node while $\Delta_{(-)}^{+}\Delta_{(+)} \neq 0$ around the antinode, as shown in Fig. 1 & 2. This means that a higher pairing temperature requires the gap function $\Delta_{(-)}^{+}\Delta_{(+)}$ should have the d-wave or similar symmetry. Here we relate the pairing temperature to the electronic structure, the electronic structure is strictly related to the crystal structure, and thus the relation between pairing temperature and crystal structure [8] could be understood. Because the d-wave symmetry is related to the anisotropy, thus that the tetragonal structure [9] corresponds to higher pairing temperature is also understood.

The electron correlation affects $T_{pair}$ through the two contradictory ways. That the electron correlation intends to increase $T_{pair}$ could be found with $V(q) + J(q)$ in Eq. (15). However, that the electron correlation intends to decrease $T_{pair}$ could also be found with the difference $E_k^{(+)} - E_k^{(-)}$ in Eq. (15): the stronger the electron correlation is, the larger the difference $E_k^{(+)} - E_k^{(-)}$ (for some wave vectors) is, and the lower the pairing temperature is. Therefore, we explained why the high-temperature superconductivity occurs near the Mott metal-insulator transition [10]. In other words, the high-temperature superconductivity occurs at the materials in which the electron correlations are moderate.

Because $E_k^{(+)} <0$ in the antinodal region, the pairs around the antinode within the energy band $E_k^{(+)}$ are not on the Fermi surface and then are not responsible for the superconductivity. However, the pairs give an energy gap, thus the energy gap is the pseudogap. To result in the superconductivity, some of the pairs must appear on the Fermi surface (this means that $E_F - \mu = E_{k_F}^{(+)}=0$ for $\Delta_{(-)}^+\Delta_{(+)} =0$ on some Fermi segments while $E_k^{(+)} \neq 0$ for $\Delta_{(-)}^+\Delta_{(+)} \neq 0$ around the Fermi segments), thus some pairs must take the place of some segments of the Fermi arc near the four nodes for $T<T_c$ or $T \to T_c - 0$.

As discussed above, to get a higher pairing temperature, $E_k^{(+)} - E_k^{(-)}$ should tend to zero in as many as possible points near or on the Fermi surface, thus $T_c<T_{ps}$ could be understood since the pairs around node will lead to $E_k^{(+)} - E_k^{(-)} \neq 0$ in more points on the Fermi surface for $T \to T_c - 0$.

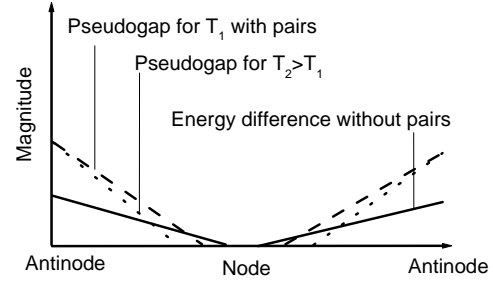

Fig.1: Schematic gap structure of high-$T$c superconductor in underdoped region. Dash line: pseudogap function for $T_1$ with pairs; dot line: pseudogap function for $T_2 >T_1$ with pairs; solid line: energy difference $E_k^{(+)} - E_k^{(-)}$ without pairs. The Fermi arc could be observed around node. Some pairs transfer toward node with the decreased temperature.

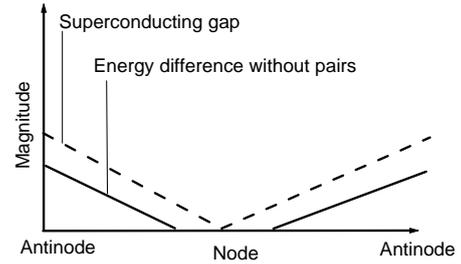

Fig.2. Schematic gap structure of high-$T$c superconductor in optimally doped region for T<$T_c$. Dash line: superconducting gap with pairs; solid line: energy difference $E_k^{(+)} - E_k^{(-)}$ without pairs. While whether superconducting gap is larger than energy difference without pairs is determined by temperature and materials.

## 4. Compare with experiment

The discussion above arrives at such a conclusion that the superconductivity requires the transfer of the pairing weight from the antinodal region to the nodal region with the change of doping or temperature. This conclusion is in agreement with the experiment [11]. That is to say, Kondo and coauthors experiment shows in fact that some pairing positions can transfer from the antinode to the node in the wave vector space, or from the region far from the Fermi level to the region near Fermi surface segments, with the change of doping or temperature. That $W_{PG}$ increases with the decreased temperature for $T^*>T>T_c$ has two possible causes. One, the number of the pairs is increased by the decreased temperature. Two, the electron excitations from the pairing region are suppressed. When some pairing space transfer toward the Fermi level for $T<T_c$, while the number of these pairs have remained almost unvaried, $W_{PG}$ should decrease while $W_{CP}$ increase, thus the almost perfect linear anti-correlation between $W_{CP}$ and $W_{PG}$ can be qualitatively understood. When some pairs are formed near the nodal region which is on the Fermi surface, the superconductivity may occur.

Moreover, other experiment data could be explained with the same viewpoint. For example, the Fermi pocket observed in experiment [12] can be explained if the effect of free electron states (the free electron states and the localized electron orbits belong to the same Hamiltonian $H_0$ as discussed in previous paper)[13], while more details have to be discussed in another paper.

## 5. Conclusion

In summary, the pairing temperature $T_{pair}$ could be explained as the pseudogap temperature $T_{ps}$ when the pairs are not on the Fermi surface, while the pairing temperature $T_{pair}$ should be explained as the superconducting critical temperature $T_c$ when some pairs are on the Fermi surface. Thus the coexistence of the pseudogap and the superconducting gap is explained when these two kinds of gaps are not all on the Fermi surface. It is also shown that the symmetry of the pseudogap and the superconducting gap are determined by the electronic structure. The high-temperature superconductivity near the Mott metal-insulator transition is also understood.